\input harvmac
\input epsf
\noblackbox
\newcount\figno
\figno=0
\def\fig#1#2#3{
\par\begingroup\parindent=0pt\leftskip=1cm\rightskip=1cm\parindent=0pt
\baselineskip=11pt
\global\advance\figno by 1
\midinsert
\epsfxsize=#3
\centerline{\epsfbox{#2}}
\vskip 12pt
\centerline{{\bf Figure \the\figno} #1}\par
\endinsert\endgroup\par}
\def\figlabel#1{\xdef#1{\the\figno}}
\def\pano{\par\noindent}
\def\smno{\smallskip\noindent}

\def\bigno{\bigskip\noindent}
\font\cmss=cmss10
\font\cmsss=cmss10 at 7pt
\def\rlx{\relax\leavevmode}
\def\inbar{\vrule height1.5ex width.4pt depth0pt}
\def\IC{\relax\,\hbox{$\inbar\kern-.3em{\rm C}$}}
\def\IR{\relax{\rm I\kern-.18em R}}
\def\IN{\relax{\rm I\kern-.18em N}}
\def\IP{\relax{\rm I\kern-.18em P}}
\def\ZZ{\rlx\leavevmode\ifmmode\mathchoice{\hbox{\cmss Z\kern-.4em Z}}
 {\hbox{\cmss Z\kern-.4em Z}}{\lower.9pt\hbox{\cmsss Z\kern-.36em Z}}
 {\lower1.2pt\hbox{\cmsss Z\kern-.36em Z}}\else{\cmss Z\kern-.4em Z}\fi}
\def\narrowplus{\kern -.04truein + \kern -.03truein}
\def\narrowminus{- \kern -.04truein}
\def\narrowminussub{\kern -.02truein - \kern -.01truein}
\def\cl{\centerline}
\def\a{\alpha}

\def\oX{\overline{X}}
\def\oY{\overline{Y}}
\def\o#1{\overline{#1}}

\def\ra{\rangle}
\def\cK{{\cal{K}}}
\def\cA{{\cal{A}}}
\def\cM{{\cal{M}}}
\def\th#1#2{\vartheta\bigl[{  #1 \atop #2} \bigr] }



\lref\rsagbi{A. Sagnotti, M. Bianchi,
{\it On the Systematics of Open String Theories},
Phys.\ Lett.\ {\bf B247} (1990) 517.}

\lref\rnonsusy{
A. Sagnotti, {\it Some Properties of Open String Theories},
 hep-th/9509080;
A. Sagnotti, {\it Surprises in Open String Perturbation Theory},
hep-th/9702093;
C. Angelantonj, {\it Non-Tachyonic Open Descendants of the 
0B String Theory}, Phys.Lett. {\bf B444} (1998) 309, hep-th/9810214;
R. Blumenhagen, A. Font and D. L\"ust, {\it Tachyon-free
Orientifolds of Type 0B Strings in Various Dimensions}, hep-th/9904069;
R. Blumenhagen, A. Kumar, {\it A Note on Orientifolds and Dualities of 
Type 0B String Theory}, hep-th/9906234.} 

\lref\rsato{S. Ishihara, H. Kataoka and H. Sato, {\it D=4, N=1, Type 
IIA Orientifolds}, hep-th/9908017.}

\lref\rgimpol{
E.G.\ Gimon and J.\ Polchinski, {\it Consistency Conditions
for Orientifolds and D-Manifolds}, Phys.\ Rev.\ {\bf D54} (1996) 1667,
hep-th/9601038.}

\lref\rgimjo{ E.G.\ Gimon and C.V.\ Johnson, {\it K3 Orientifolds},
Nucl.\ Phys.\ {\bf B477} (1996) 715, hep-th/9604129.}

\lref\rberk{ M. Berkooz and  G. Leigh, {\it A D=4 N=1 Orbifold of Type I 
Strings}, Nucl.Phys. {\bf B483} (1997) 187, hep-th/9605049;
M. Berkooz, R.G. Leigh, J. Polchinski, J. H. Schwarz, N.  Seiberg, E. Witten,
{\it Anomalies, Dualities, and Topology of D=6 N=1 Superstring Vacua},
Nucl.Phys. {\bf B475} (1996) 115, hep-th/9605184.}

\lref\rkakusix{Z. Kakushadze, G. Shiu and S.-H. H. Tye, {\it 
Type IIB Orientifolds with NS-NS Antisymmetric Tensor Backgrounds},
Phys.Rev. {\bf D58} (1998) 086001, hep-th/9803141;
A. Buchel, G. Shiu, S.-H. H. Tye, {\it Anomaly Cancelations in Orientifolds 
with Quantized B Flux}, hep-th/9907203.}

\lref\rkakufour{Z. Kakushadze, {\it Aspects of N=1 Type I-Heterotic Duality 
in 
Four Dimensions}, Nucl.Phys. {\bf B512} (1998) 221, hep-th/9704059;
Z. Kakushadze and G. Shiu, {\it A Chiral N=1 Type I Vacuum in Four 
Dimensions and Its Heterotic Dual}, Phys.Rev. {\bf D56} (1997) 3686, 
hep-th/9705163; Z. Kakushadze and G. Shiu, {\it 4-D Chiral N=1 Type I 
Vacua With And Without D5-Branes}, Nucl.Phys. {\bf B520} (1998) 75,
 hep-th/9706051; Z. Kakushadze, {\it On Four Dimensional N=1 Type I 
Compactifications}, Nucl.Phys. {\bf B535} (1998) 311, hep-th/9806008.}

\lref\rkakunp{ Z. Kakushadze, G. Shiu and  S.-H. H. Tye, {\it
   Type IIB Orientifolds, F-theory, Type I Strings on Orbifolds and 
    Type I - Heterotic Duality}, Nucl. Phys. {\bf B533} (1998) 25,
    hep-th/9804092;
   Z. Kakushadze, {\it Non-perturbative Orientifolds}, hep-th/9904007;
  Z. Kakushadze, {\it Type I on (Generalized) Voisin-Borcea Orbifolds and 
  Non-perturbative Orientifolds}, hep-th/9904211;
   Z. Kakushadze, {\it Non-perturbative K3 Orientifolds with NS-NS B-flux},
  hep-th/9905033. }

\lref\rblum{J.D. Blum and A. Zaffaroni, {\it An Orientifold from F Theory},
Phys.Lett. {\bf B387} (1996) 71, hep-th/9607019;
J.D. Blum, {\it F Theory Orientifolds, M Theory Orientifolds and 
Twisted Strings}, Nucl.Phys. {\bf B486} (1997) 34, hep-th/9608053.}

\lref\rbg{R. Blumenhagen and L. G\"orlich, {\it Orientifolds of 
Non-Supersymmetric Asymmetric Orbifolds}, Nucl.Phys. {\bf B551} (1999) 601,
hep-th/9812158.}

\lref\rfut{R. Blumenhagen, L. G\"orlich and B. K\"ors, {\it in 
preparation} .}

\lref\rfutb{R. Blumenhagen and A. Kumar, {\it in 
preparation} .}

\lref\rbw{R. Blumenhagen and A. Wi\ss kirchen, {\it Spectra of 4D, N=1 
Type I String Vacua on Non-Toroidal CY Threefolds}, 
Phys.Lett. {\bf B438} (1998) 52, hep-th/9806131.} 

\lref\rzwart{G. Zwart, {\it Four-dimensional N=1 $Z_N\times Z_M$ 
Orientifolds}, Nucl.Phys. {\bf B526} (1998) 378, hep-th/9708040.}

\lref\rang{C. Angelantonj, M. Bianchi, G.Pradisi, A. Sagnotti and Y. Stanev,
{\it Chiral Asymmetry in Four-Dimensional Open-String Vacua}, 
Phys.Lett. {\bf B385} (1996) 96, hep-th/9606169.}

\lref\rgep{C. Angelantonj, M. Bianchi, G. Pradisi, A. Sagnotti and
Ya. S. Stanev, {\it Comments on Gepner Models and Type I Vacua in String
Theory}, Phys.Lett. {\bf B387} (1996) 743, hep-th/9607229.} 

\lref\rgopa{R. Gopakumar and S. Mukhi, {\it Orbifold and Orientifold 
Compactifications of F-Theory and M-Theory to Six and Four Dimensions},
Nucl.Phys. {\bf B479} (1996) 260, hep-th/9607057.} 

\lref\rdabol{ A. Dabholkar and J. Park, {\it An Orientifold of Type 
        IIB theory on K3}, Nucl. Phys. {\bf B472} (1996) 207, 
        hep-th/9602030;
        {\it Strings on Orientifolds}, Nucl. Phys. {\bf B477}
        (1996) 701, hep-th/9604178.}

\lref\rbfield{  M. Bianchi, G. Pradisi and A. Sagnotti, 
{\it Toroidal Compactification and Symmetry Breaking in Open-String 
    Theories}, 
    Nucl.Phys. {\bf B376} (1992) 365;
M.\ Bianchi, {\it A Note on Toroidal Compactifications of the Type I 
 Superstring and Other Superstring Vacuum Configurations with 16 
  Supercharges}, Nucl. Phys. {\bf B528} (1998) 73, hep-th/9711201;
E.\ Witten, {\it Toroidal Compactification Without Vector Structure},
     JHEP {\bf 9802} (1998) 006, hep-th/9712028;
C. Angelantonj, {\it Comments on Open-String Orbifolds with a Non-Vanishing
$B_{ab}$}, hep-th/9908064.}

\lref\riban{G. Aldazabal, A. Font, L. E. Ib\'a\~nez and G. Violero, {\it
   D=4,N=1, Type IIB Orientifolds},   Nucl.Phys. {\bf B536} (1998) 29,
        hep-th/9804026; L. E. Ib\'a\~nez, R. Rabad\'an and A. M. Uranga, 
{\it Anomalous U(1)'s in Type I and Type IIB D=4, N=1 String Vacua},
Nucl.Phys. {\bf B542} (1999) 112, hep-th/9808139;
G. Aldazabal, D. Badagnani, L. E. Ib\'a\~nez and A. M. Uranga, {\it
Tadpole Versus Anomaly Cancellation in D=4,6 Compact IIB Orientifolds},
JHEP {\bf 9906} (1999) 031, hep-th/9904071}    

\lref\rangles{M. Berkooz, M. R. Douglas and  R.G. Leigh, {\it Branes 
Intersecting at Angles}, Nucl.Phys. {\bf B480} (1996) 265, hep-th/9606139;
H. Arfaei and M. M. Sheikh Jabbari, {\it Different D-brane Interactions},
Phys.Lett. {\bf B394} (1997) 288, hep-th/9608167;
J. C. Breckenridge, G. Michaud and R. C. Myers, {\it New angles on D-branes},
Phys.Rev. {\bf D56} (1997) 5172, hep-th/9703041;
M. M. Sheikh Jabbari, {\it Classification of Different Branes at Angles},
Phys.Lett. {\bf B420} (1998) 279, hep-th/9710121.}

\lref\rtev{I. Antoniadis, {\it A Possible New Dimension at a Few TeV},
Phys.Lett. {\bf B246} (1990) 377;
J. Lykken, {\it Weak Scale Superstrings}, 
Phys.Rev. {\bf D54} (1996) 3693, hep-th/9603133;
N. Arkani-Hamed, S. Dimopoulos and G. Dvali, 
{\it The Hierarchy Problem and New Dimensions at a Millimeter}, 
Phys.Lett. {\bf B429} (1998) 263, hep-ph/9803315;
N. Arkani-Hamed, S. Dimopoulos and G. Dvali,
{\it Phenomenology, Astrophysics and Cosmology of Theories with Sub-Millimeter
       Dimensions and TeV Scale Quantum Gravity},
Phys.Rev. {\bf D59} (1999) 086004, hep-ph/9807344;
I. Antoniadis, N. Arkani-Hamed, S. Dimopoulos, G. Dvali, {\it
New Dimensions at a Millimeter to a Fermi and Superstrings at a TeV},
Phys.Lett. {\bf B436} (1998) 257, hep-ph/9804398;
 G. Shiu, S.-H. H. Tye, {\it TeV Scale Superstring and Extra Dimensions},
Phys.Rev. {\bf D58} (1998) 106007, hep-th/9805157.}

\Title{\vbox{\hbox{hep--th/9908130}
 \hbox{HUB--EP--99/45}}}
{\vbox{\centerline{Supersymmetric Orientifolds in 6D}
\bigskip\centerline{with D-Branes at Angles}
}}
\centerline{Ralph Blumenhagen${}^1$, Lars G\"orlich${}^2$  and 
Boris K\"ors${}^3$}
\bigskip
\centerline{\it ${}^{1,2,3}$ Humboldt-Universit\"at zu Berlin, Institut f\"ur 
Physik,}
\centerline{\it \   Invalidenstrasse 110, 10115 Berlin, Germany }
\smallskip
\bigskip
\centerline{\bf Abstract}
\noindent
We study  a new class of N=1 supersymmetric 
orientifolds in six space-time dimensions. The world-sheet parity
transformation is combined with a permutation of the internal complex
coordinates. In contrast to ordinary orientifolds the twisted sectors 
contribute to the Klein bottle amplitude leading to new tadpoles to 
be cancelled by twisted open string sectors. They arise from 
open strings stretched  between D7-branes intersecting at non-trivial angles.
We study in detail the $\ZZ_3$, $\ZZ_4$ and $\ZZ_6$ permutational
orientifolds obtaining in all cases anomaly free massless spectra. 
\footnote{}
{\pano
${}^1$ e--mail:\ blumenha@physik.hu-berlin.de
\pano
${}^2$ e--mail:\ goerlich@physik.hu-berlin.de
\pano
${}^2$ e--mail:\ koers@physik.hu-berlin.de
\pano}
\Date{08/99}

\newsec{Introduction}

In recent years we have seen the main activity in
string model building shifting  from
the foremost promising class of heterotic Calabi-Yau compactifications
to compact Type I and  orientifold models. 
The latter class generically contains D-branes in the background supporting
the gauge sector of the low energy theory, whereas gravity propagates 
in the ten-dimensional bulk. Therefore, such models became
quite attractive in recent attempts to establish a unification
scenario with the string scale as low as 1TeV \rtev. 

In contrast to the millions  of consistent models partly classified
in the Calabi-Yau setting, so far we know comparatively few examples of
orientifold models. This is  mainly due to the fact that  the only
consistency condition we know of, namely tadpole cancellation,  requires
the complete knowledge of the one-loop partition function.
Therefore we find ourselves restricted to toroidal orbifolds and Gepner 
models which were discussed in  
six  flat dimensions in 
\refs{\rsagbi\rgimpol\rdabol\rgimjo\rblum\rgep-\rkakusix} and in four 
dimensions in \refs{\rberk\rang\rkakufour\rzwart\riban\rgopa-\rbw}.

In this paper we study a new class of orientifolds of Type IIB 
where the world-sheet parity transformation
is combined with an ${\cal S}_2$ 
permutation of the internal complex coordinates. In real coordinates the
permutation is nothing else than a reflection of some of the internal
coordinates, not preserving the complex structure.  
As a consequence, in these permutational orientifolds  the loop channel
twisted sector Klein bottle amplitudes 
are non-vanishing. It turns out that in  tree channel only
untwisted and $\ZZ_2$ twisted sectors propagate between the two 
cross-caps.  
In order to cancel tadpoles it is necessary to introduce D7-branes
intersecting at non-trivial angles providing in open string loop channel 
sectors, which can be regarded as twisted open string sectors.
As we will 
show, these new sectors automatically arise due to the geometric $\ZZ_N$
symmetry which simply rotates the D7-branes by an angle $\phi=2\pi/N$. 
Inclusion of these twisted  open string sectors indeed allows us
to cancel the Klein bottle tadpoles and finally leads to massless closed
and open string spectra satisfying the N=1 anomaly constraint.

In a different class of models \rkakunp\ the request for 
twisted open string sectors was expressed, as well, 
however in the models discussed there also all twisted sectors
appear in the tree channel Klein bottle amplitude.
Thus, there are more tadpoles to cancel, which was argued to
be impossible just by perturbative open string sectors.
Let us emphasize that the models presented here and the models in \rkakunp\
are not related by T-duality, as T-duality maps the left-right symmetric
$\ZZ_N$ action to an asymmetric one. 

Moreover, since in the permutational orientifolds all twisted sectors
are left invariant,  the non-supersymmetric Type 0B generalization \rfutb\ 
could lead to new tachyon-free orientifolds of the type discussed in
\rnonsusy. 

This paper is organized as follows. In section two we will define
our models and present the computation of the $\ZZ_3$ orientifold
in very much detail. In section three we will discuss the $\ZZ_4$ and $\ZZ_6$ 
orientifolds, thereby focusing mostly on the new issues 
compared to the $\ZZ_3$ case. Finally, we will end with some
conclusions.

\newsec{The $\ZZ_3$ permutational orientifold}

The six dimensional models studied in the course of this paper are
defined as follows. We compactify Type IIB on the four torus $T^4$
with two complex coordinates $X=x_8+ix_9$ and $Y=x_6+i x_7$. 
Then we take the orientifold by the group $G+\Omega R G$, where 
the space-time symmetry group is $G=\ZZ_N$ with one of the
four choices $N\in\{2,3,4,6\}$ allowing for a crystallographic action. 
The supersymmetry preserving action of $\ZZ_N$
on the two complex coordinates is given by
\eqn\chapaa{  \Theta:\cases{ X\to e^{2\pi i{1\over N}}X \cr 
              Y\to e^{-2\pi i{1\over N}}Y\cr} .}
In the orientifold the world-sheet parity transformation is always paired
with the following $\ZZ_2$ operation
\eqn\chapab{  R:\cases{ X\to \oX \cr Y\to \oY\cr } .}
Thus, from the perspective of the complex coordinates it is nothing else
than a permutation, whereas from the perspective of the real coordinates 
$x_i$ it is the reflection in the $x_7$ and $x_9$ direction.
We also have to specify the action of $R$ on the Ramond sector ground
states $R:|s_1\, s_2\, s_3\, s_4\ra\to |s_1\, s_2\, -s_3\, -s_4\ra$.
Note, that the combination $\Omega R$ is related to $\Omega$ by
T-duality in the $x_7$ and $x_9$ direction. However, under this T-duality
the geometric $\ZZ_N$ operation $\Theta$ is mapped to $T\Theta T^{-1}=
\hat\Theta$ acting in an asymmetric way on the  left and right moving
components of the fields $X$ and $Y$ 
\eqn\chapabb{  \hat\Theta:\cases{ X_L\to e^{2\pi i{1\over N}}X_L, \quad 
              X_R\to e^{-2\pi i{1\over N}}X_R \cr
               Y_L\to e^{-2\pi i{1\over N}}Y_L, \quad 
              Y_R\to e^{2\pi i{1\over N}}Y_R \cr}.}
Thus, the orientifolds we are going to discuss are not related via
T-duality to the ordinary orientifolds. 

To our knowledge permutational orientifolds were discussed for
the first time in \rbg\ in the context of non-supersymmetric string theory.  
The combination of $\Omega$ with $R$ has dramatic consequences for the
computation of the tadpole cancellation conditions as compared to usual 
orientifold models.
First of all, we observe that $R$ does not commute with $\ZZ_N$, instead
one encounters the relation 
\eqn\chapaab{  R \Theta^k = \Theta^{N-k} R, }
where $\Theta$ denotes the generator of $\ZZ_N$. 
As a consequence, using the geometry of the Klein bottle as shown
in figure 1 one obtains  for the twist 
$g=(\Omega R \Theta^k)^2=(\Omega R \Theta^l)^2=1$, 
so that in the tree channel only untwisted closed string states propagate
along the tube. Therefore, we expect to find only untwisted tadpoles. 
\fig{}{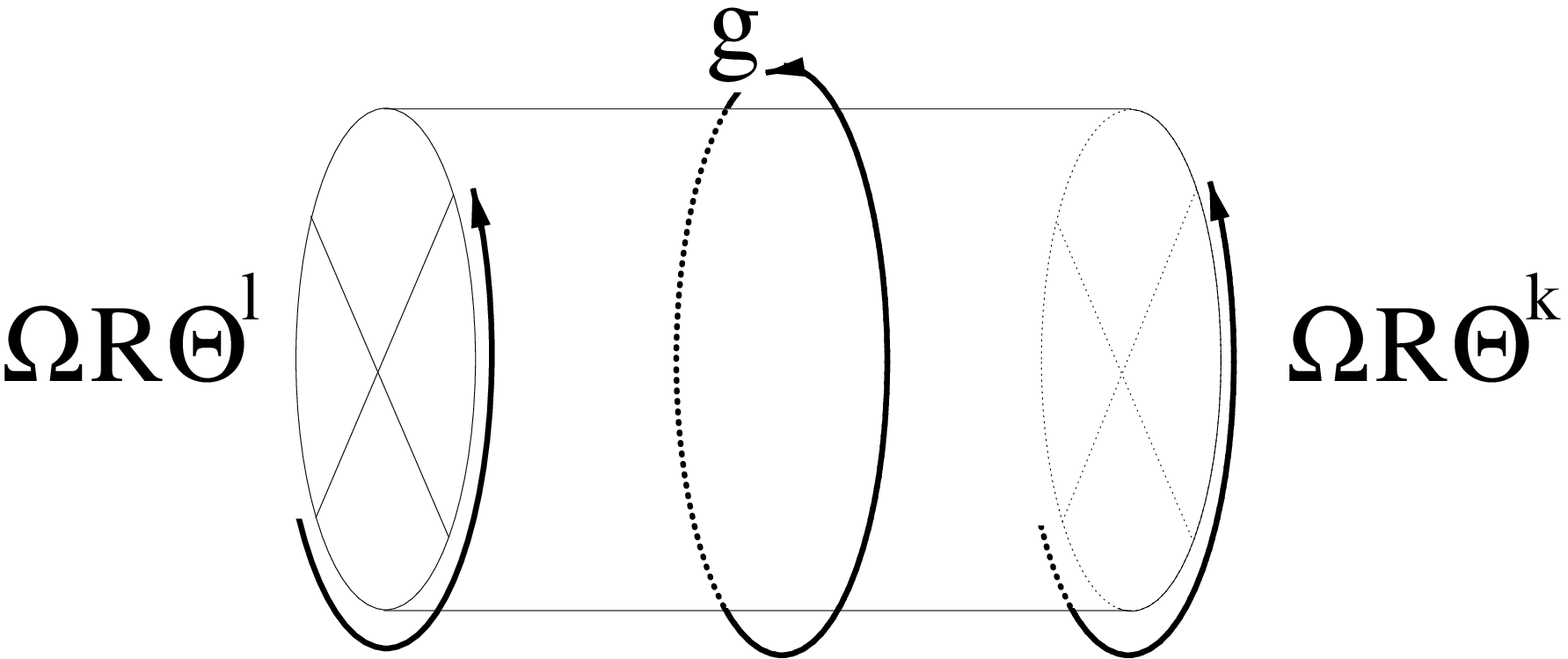}{7truecm}

Moreover, in all orientifold models studied so far, 
the action of the world-sheet  parity on 
the twisted sector ground states was such, that the sector twisted
by $\Theta^k$ was mapped to the sector twisted by $\Theta^{N-k}$.
Thus, only the untwisted
and some $\ZZ_2$ twisted sectors gave a non-vanishing contribution 
to the Klein bottle amplitude, which were cancelled by annulus and 
M\"obius contributions of D9- and D5-branes in most cases. 
Now, the permutation $R$
by itself also exchanges the $k$ twisted sector with the $N-k$ twisted
sector, as can be seen directly from the definition of the twisted sector. 
In the $k$ twisted sector the field $X$ has the following monodromy
\eqn\chapac{ X(\sigma+2\pi,\tau)=e^{2\pi i{k\over N}} X(\sigma,\tau),\quad
  \quad \oX(\sigma+2\pi,\tau)=e^{-2\pi i{k\over N}} \oX(\sigma,\tau) ,}
which this mapped under the action of $R$ to
\eqn\chapad{ \oX(\sigma+2\pi,\tau)=e^{2\pi i{k\over N}} \oX(\sigma,\tau),\quad
  \quad X(\sigma+2\pi,\tau)=e^{-2\pi i{k\over N}} X(\sigma,\tau) .}
Thus, after applying $R$ the field $X$ lies  in the $N-k$ twisted sector. 
Therefore, the combined operation $\Omega R$ leaves the twisted sector 
invariant and one expects non-vanishing contributions to the Klein-bottle
amplitude from all $N-1$ twisted sectors\footnote{$^1$}{This essential
point was not realized  in \rsato\ and renders the whole computation
performed there fairly questionable.}.
Note, that the $k$ twisted sectors of the asymmetric 
$\widehat{\ZZ}_3$ orbifold are left invariant under $\Omega$ and are mapped
to the $N-k$ twisted sectors under $\Omega R$. This   confirms the T-duality
relation of these models to ordinary orientifolds. 

We are now facing the problem 
of how to introduce ``twisted'' open string sectors with the ability to
cancel these tadpoles arising in the twisted Klein bottle amplitudes. 
In usual orientifolds 
open strings stretched between the same type of D-branes can be regarded as
untwisted open string sectors  and open strings stretched between a D9- and
a D5-brane can be regarded as $\ZZ_2$ twisted open string sectors.

The problem of twisted open string sectors was also raised in a 
slightly different class of orientifolds
in \rkakunp, where  it was simply enforced by hand that $\Omega$ did
not exchange the twisted sectors. 
The conclusion there was, that
these ``twisted''  open string sectors are non-perturbative in nature and
could  be detected by using  Type I - heterotic  duality. We would like
to make the following point clear. We are not 
claiming that these ``non-perturbative'' states found there 
do allow for a perturbative description. 
What we will show is, that in a different class of
orientifolds twisted open string sectors 
can indeed be described in a perturbative way, namely
by open strings stretched between D7-branes intersecting at non-trivial
angles. 
 
In the rest of this section we will construct
the $\ZZ_3$ permutational orientifold in some detail revealing 
how nicely everything fits together. 
Before that  let us remark, that the $\ZZ_2$ permutational orientifold is 
isomorphic
to the ordinary one, as in that case the action of $\ZZ_2$ is the same
on all four compact coordinates and it is a matter of taste which two
are named X and Y. Thus, in that case one simply gets the T-dual of
the Gimon-Polchinski model \rgimpol\ containing instead of D9- and D5-branes
two different sorts of D7-branes. 
The first non-trivial example is the $\ZZ_3$ orientifold.

\subsec{Klein bottle amplitude}

Before we can compute the Klein bottle amplitude we need to review
some facts about the two dimensional $\ZZ_3$ lattice. 
In both the (67) and the (89) plane we choose the elementary cell
of the $T^2$ torus as shown in figure 2.
\fig{}{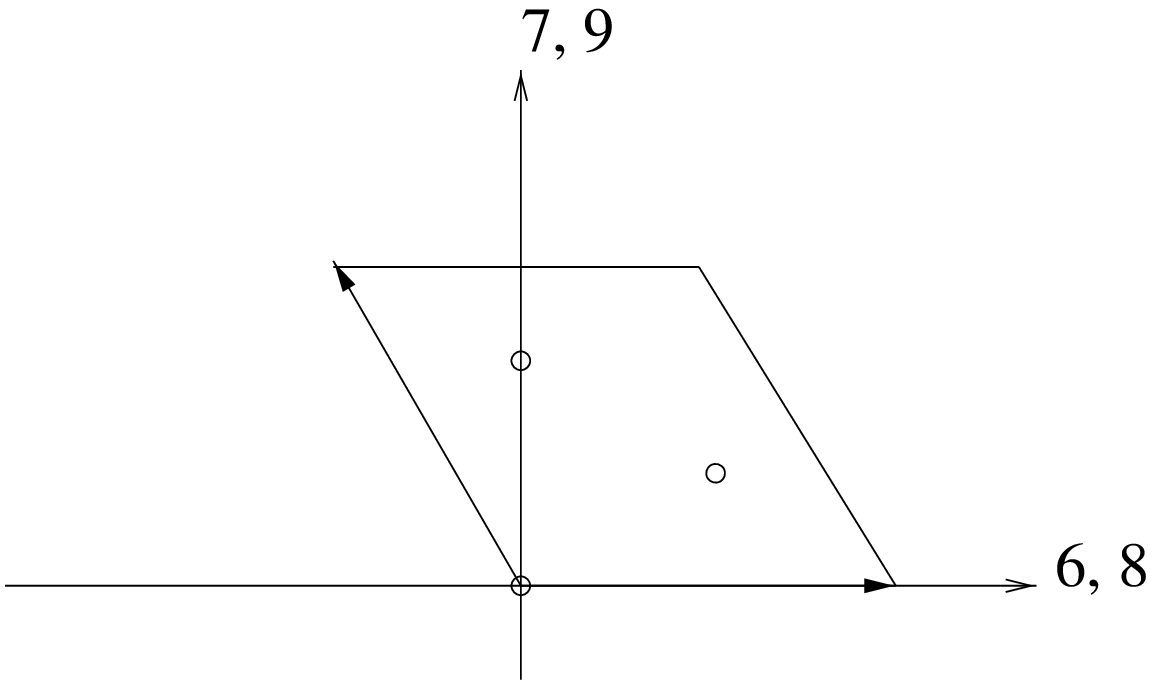}{7truecm}
\noindent
The left-right moving momenta are given by
\eqn\chapae{ p_{L,R}=P^I\pm {1\over 2} L^I }
with Kaluza-Klein (KK) momentum and winding given by 
\eqn\chapaf{ P^I={\sqrt{2}\over r}\left( m_1 \vec{e}_1^{\, *} +  
  m_2 \vec{e}_2^{\, *}  \right),\quad\quad 
     L^I={r\over \sqrt{2}}\left( n_1 \vec{e}_1 +  n_2 \vec{e}_2 \right). }
Here we have chosen $\alpha'=2$. 
The basis and the dual basis  are 
\eqn\chapag{\eqalign{  &\vec{e}_1=(\sqrt 2,0),\quad\quad 
                        \vec{e}_2=(-{1\over \sqrt 2},\sqrt{3\over 2}) \cr
                       &\vec{e}_1^{\, *}=({1\over \sqrt 2},{1\over \sqrt 6}),
                      \quad\quad 
                       \vec{e}_2^{\, *}=(0,\sqrt{2\over 3}). \cr}}
In figure 2 we have also denoted the three fixed points of the $\ZZ_3$ action
on each $T^2$.
Note, that under the permutation $R$ the origin is invariant whereas the
other two fixed points get exchanged.  
Now we have to compute the  Klein bottle amplitude
\eqn\chapah{ \cK= 8\, c \, \int_0^\infty {dt\over t^4}\, 
                  {\rm Tr}_{U+T} \left[
                  {\Omega R\over 2}\, 
            \left( {1+\Theta+\Theta^2\over 3}\right)\, P_{GSO}\, 
                   e^{-2\pi t(L_0+\o{L}_0)} \right] ,}
where the momentum integration in the non-compact directions has
already been carried out and $c = V_{6}/(8\pi^2 \a')^3$.
Let us first discuss the untwisted sector.
\smno
a.) {\it Untwisted sector}
\smno
Unlike the purely $\Omega$ orientifold where left-right combinations
of  states of the schematic form
$X \tilde{X}$  and $Y \tilde{Y}$ were contributing in the Klein bottle 
amplitude,
in our case the states $X\widetilde{\overline{X}}$ and 
$Y\widetilde{\overline{Y}}$ are relevant. 
Since $\Theta$ acts on these states as
\eqn\chapai{  \Theta(X\widetilde{\overline{X}} )=
                    X\widetilde{\overline{X}},\quad\quad
               \Theta(Y\widetilde{\overline{Y}} )=
                    Y \widetilde{\overline{Y}} }
i.e. without any phase factor,  each term in the sum 
$(1+\Theta+\Theta^2)/3$ in \chapah\
yields the same result in the oscillator part of the trace. 

This is also true for the lattice part due to its $\ZZ_3$ symmetry. 
Note that $R$ is the reflection at the axes defined by $\vec e_1$, 
$R\Theta$  is the reflection at $\vec e_2$ and $R\Theta^2$  
is the reflection at $\vec e_1+\vec e_2$. 
Under the action of $\Omega R$ only KK momentum in the $x_6$ and $x_8$
direction and winding in the $x_7$ and $x_9$ direction survives. 
Using the expression in \chapaf\ and \chapag, one finds that
$P^6=(2m)/r$, $L^7=\sqrt{3} nr$ and similarly for the second torus.
It is now straightforward to compute the untwisted Klein bottle amplitude
\eqn\chapaj{ \cK_U=c\, (1-1)\, \int_0^\infty {dt\over t^4}\, 2\,
          {\th{0}{1/2}^4 \over
        \eta^{12}} \left( \sum_m e^{-\pi t {4m^2\over \rho}} \right)^2
                   \left( \sum_n e^{-\pi t {n^2 \rho 3}} \right)^2 ,}
where the argument of the $\vartheta$-function is exp$(-4\pi t)$ and
as in \rgimpol\ $\rho=r^2/\alpha'$.
\smno
b.) {\it Twisted sectors}
\smno
In the $\ZZ_3$ twisted sector the lattice part is trivial and the action
of $\Omega R$ on the oscillator modes is analogous to \chapai. 
We have to sum over all nine fixed points in each twisted sector, 
however only one of them is
invariant under the action of $\Omega R$. Thus, the $\Theta$ twisted sector 
contribution to the Klein bottle amplitude is 
\eqn\chapak{ \cK_\Theta= c\, (1-1)\,
         \int_0^\infty {dt\over t^4}\, 
      2\, {\th{0}{1/2}^2\, \th{1/3}{1/2}\, \th{-1/3}{1/2} \over
        \eta^{6} \,  \th{-1/6}{1/2}\, \th{1/6}{1/2} } .}
\smno
The $\Theta^2$ twisted sector 
contribution to the Klein bottle amplitude can be obtained 
simply by exchanging $\th{a}{1/2}$ with $\th{-a}{1/2}$ in \chapak\ and
actually leads to the same partition function. 
Having everything expressed in terms of $\vartheta$-functions
the transformation into tree-channel is straightforward and yields for
the complete Klein bottle amplitude
\eqn\chapal{\eqalign{ \tilde{\cK}=c\, (1-1)\, &\int_0^\infty dl\, 
 {2^5\over 3}\,  \Biggl[ {\th{1/2}{0}^4 \over
        \eta^{12}} \left( \sum_m e^{-\pi l {m^2 \rho}} \right)^2
                \left( \sum_n e^{-\pi l {4n^2 \over 3 \rho }} \right)^2 +\cr
     &\cr
      &3{\th{1/2}{0}^2\, \th{1/2}{1/3}\, \th{1/2}{-1/3} \over
        \eta^{6}\,   \th{1/2}{-1/6}\, \th{1/2}{1/6}} +
      3{\th{1/2}{0}^2\, \th{1/2}{-1/3}\, \th{1/2}{1/3} \over
        \eta^{6}\,   \th{1/2}{1/6}\, \th{1/2}{-1/6}} \Biggr] \cr }}
with the argument exp$(-4\pi l)$. 
As expected we only find untwisted tadpoles and the numerical factors
in front of the $\vartheta$-functions in \chapal\ are exactly 
$4\sin^2(\pi k/N)$. 
This guarantees that only $\ZZ_3$ invariant states from 
the untwisted closed string sector 
propagate along the tube of the tree channel Klein bottle. 
In other words, the three different terms
in \chapal\ really constitute the complete projector $(1+\Theta+\Theta^2)/3$
acting on the untwisted states in tree channel. Note, from our loop channel 
computation 
this matching appears to be some nice conspiracy 
between the numerical lattice factors from the
untwisted sector and the number of invariant fixed points in the twisted
sectors. In the open string sector we 
will promote this ``completion of the projector in tree-channel'' to our
guiding principle in determining the relative normalizations of the
different open string sectors. 

\subsec{Annulus amplitude}

Since via T-duality in the $x_7$ and $x_9$ directions $\Omega R$ is
related to $\Omega$, we expect that we need D7-branes to cancel the 
Klein-bottle tadpoles. Moreover, in the four internal coordinates these
D7-branes should extend in the $x_6$ and $x_8$ direction and fill the
complete six-dimensional uncompactified space-time. 
Let us assume first that we stuck together M D7-branes on the $x_7=x_9=0$
line fixed under $R$.  
The entire 
arrangement of D7-branes has to satisfy the geometric $\ZZ_3$ symmetry 
we would like
to mod out. Therefore, we are forced to introduce rotated D7-branes,
as well. More concretely, in both the (67) and the (89) plane we get
the $\Phi_1=2\pi/3$ and $\Phi_2=4\pi/3$ rotated D7-branes as shown in 
figure 3.
\fig{}{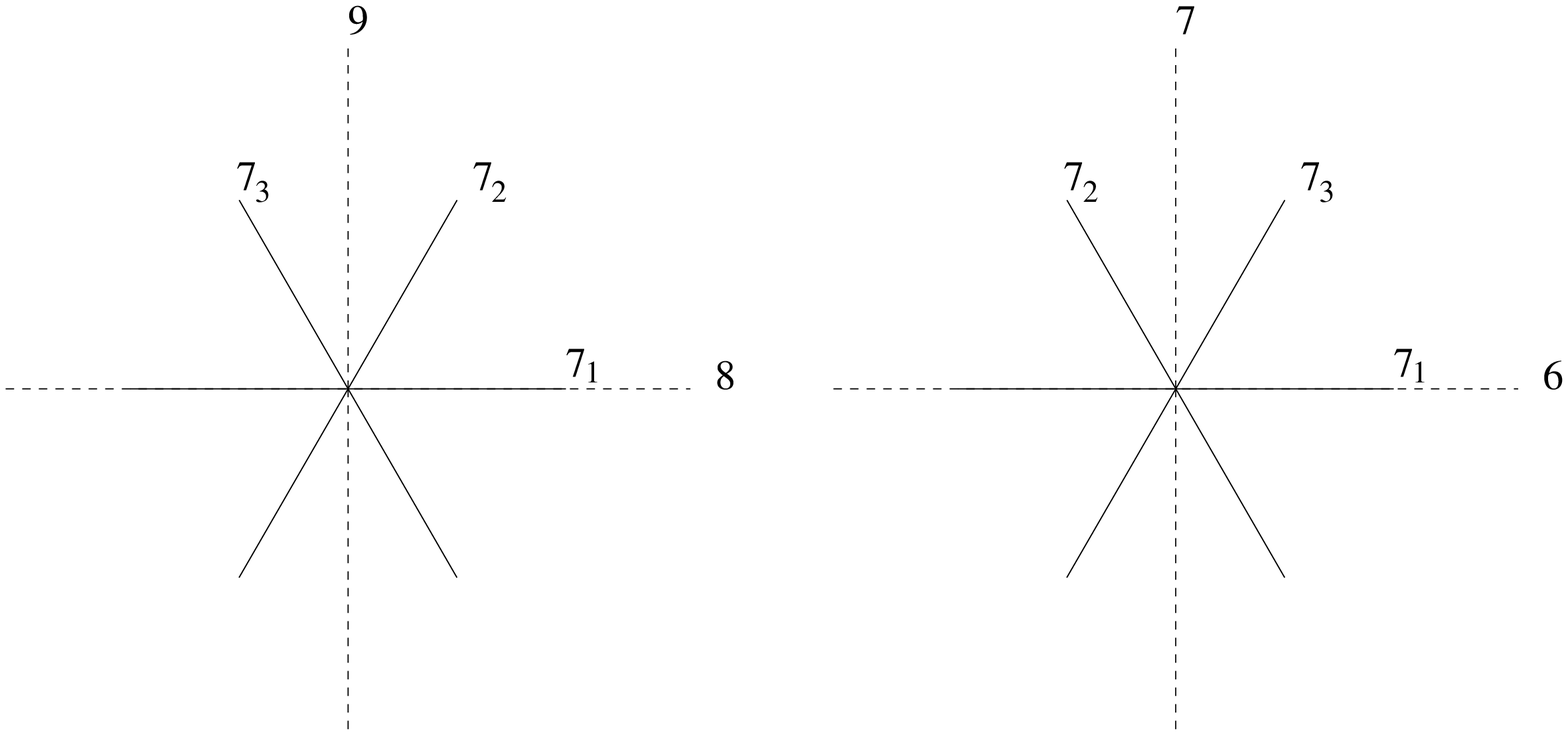}{15truecm}
\noindent
Thus, we are automatically forced to introduce D7-branes intersecting 
at non-trivial
angles which were discussed before in various papers dealing with 
supersymmetry preserving brane configurations \rangles.
Due to the analysis in \rangles\ the rotated D7-branes depicted in figure 3 
have eight unbroken supersymmetries, exactly what is needed for N=1 in
six dimensions. 
In order to derive the mode expansion for open strings connecting D-branes
at angles in the (67) and (89) plane let us restrict first to the (89) plane
and assume for simplicity that one D-brane is stretching in the $x_8$ 
direction. 
An open string ending on this D-brane satisfies the following boundary
conditions
\eqn\chapbca{ {\partial\over \partial \sigma} X^8(0,\tau)=0,
               \quad\quad
                X^9(0,\tau)=0. }
The other end of the open string ends on a different D-brane
which is rotated by an angle $\Phi=\pi\theta$ against the former
D-brane. Therefore, the boundary conditions at $\sigma=\pi$
have the following form
\eqn\chapbcb{\eqalign{   \cos(\pi\theta)\, {\partial\over \partial \sigma} 
              X^8(\pi,\tau)+\sin(\pi\theta)\,
             {\partial\over \partial \sigma} 
              X^9(\pi,\tau)&=0 \cr
            -\sin(\pi\theta)\, 
              X^8(\pi,\tau)+\cos(\pi\theta)\, 
              X^9(\pi,\tau)&=0 .\cr} } 
The solution to these two equation is
\eqn\chapbcc{\eqalign{ &X_8=\sum_{m\in\ZZ+\theta} {1\over m} \alpha_m\,
               e^{-im\tau}\, \cos(m\sigma) +
               \sum_{n\in\ZZ-\theta} {1\over n} \hat{\alpha}_n\,
               e^{-in\tau}\, \cos(n\sigma)  \cr
              &X_9=\sum_{m\in\ZZ+\theta} {1\over m} \alpha_m\,
               e^{-im\tau}\, \sin(m\sigma) -
               \sum_{n\in\ZZ-\theta} {1\over n} \hat{\alpha}_n\,
               e^{-in\tau}\, \sin(n\sigma)  \cr }}
showing that one indeed obtains something deserving the name
twisted open string sector.
The calculation for the world sheet fermions and for the 
(67) plane is completely analogous. 
In the following we will denote the number of each type of D7-brane as $M$.
The annulus amplitude 
\eqn\chapam{ \cA= c \, \int_0^\infty {dt\over t^4}\, 
                  {\rm Tr}_{open} \left[
                  {1\over 2}\, 
            \left( {1+\Theta+\Theta^2\over 3}\right)\, P_{GSO}\, 
                   e^{-2\pi t L_0} \right] }
has to be computed for all possible open strings connecting the three
kinds of D7-branes. 
\smno
a.) {\it Untwisted sector}
\smno
By untwisted open string sector we mean open strings stretched between
two D-branes of the same kind. Since the $\ZZ_3$ action rotates the 
D7-branes, 
it is clear that the $\Theta$ and $\Theta^2$ insertions in the trace in
\chapam\ vanish identically. The oscillator contribution to the
trace is as usual and, due to the $\ZZ_3$ symmetry, the lattice contribution
is the same for each type of D7-brane. More concretely, the KK momentum
in the $x_6$ and $x_8$ direction is quantized as $P=m/r$ and for
the winding in the $x_7$ and $x_9$ direction we find $L=nr\sqrt{3}/2$.
Thus, the complete untwisted annulus amplitude is 
\eqn\chapan{ \cA_U=c\, (1-1)\, \int_0^\infty {dt\over t^4}\, {M^2\over 4}\,
          {\th{0}{1/2}^4 \over
        \eta^{12}} \left( \sum_m e^{-2\pi t {m^2\over \rho}} \right)^2
                 \left( \sum_n e^{-2\pi t {n^2 \rho {3\over 4}}} \right)^2 }
with the argument exp$(-2\pi t)$.
\smno
b.) {\it Twisted sectors}
\smno
Under twisted open string sectors we understand  open strings 
stretched between different types of D7-branes.  
Since the different D7-branes  do not share any direction, 
the contribution from the lattice
to the annulus amplitude is trivial and one easily gets
\eqn\chapako{\eqalign{ \cA_\Theta&=\sum_{i=1}^3 A_{7_i 7_{i+1}}
            = \kappa_1 \  c\, (1-1)\, \int_0^\infty {dt\over t^4}\, 
      {M^2\over 4}\, {\th{0}{1/2}^2\, \th{1/3}{1/2}\, \th{-1/3}{1/2} \over
        \eta^{6} \,  \th{-1/6}{1/2}\, \th{1/6}{1/2} }, \cr
        A_{\Theta^2}&=\sum_{i=1}^3 A_{7_i 7_{i-1}}
            = \kappa_2 \  c\, (1-1)\, \int_0^\infty {dt\over t^4}\, 
      {M^2\over 4}\, {\th{0}{1/2}^2\, \th{-1/3}{1/2}\, \th{1/3}{1/2} \over
        \eta^{6} \,  \th{1/6}{1/2}\, \th{-1/6}{1/2} }, \cr}}
where in view of the more complicated $\ZZ_4$ and $\ZZ_6$ examples we 
have introduced normalization constants $\kappa_{1,2}$ 
which will be determined
by requiring that the tree channel amplitude contains the complete
$\ZZ_3$ projector. Transformation to tree channel yields
\eqn\chapap{\eqalign{ \tilde{\cA}=c\, (1-1)\, &\int_0^\infty dl\, 
 {M^2\over 6}\,  \Biggl[ {\th{1/2}{0}^4 \over
        \eta^{12}} \left( \sum_m e^{-\pi l {m^2 \rho}} \right)^2
                \left( \sum_n e^{-\pi l {4n^2 \over 3 \rho }} \right)^2 +\cr
     &\cr
      &3\, \kappa_1\, {\th{1/2}{0}^2\, \th{1/2}{1/3}\, \th{1/2}{-1/3} \over
        \eta^{6}\,   \th{1/2}{-1/6}\, \th{1/2}{1/6}} +
      3\, \kappa_2\,{\th{1/2}{0}^2\, \th{1/2}{-1/3}\, \th{1/2}{1/3} \over
        \eta^{6}\,   \th{1/2}{1/6}\, \th{1/2}{-1/6}} \Biggr] \cr }}
with the argument exp$(-4\pi l)$. With the choice $\kappa_1=\kappa_2=1$,
up to a numerical factor this is the same
as the Klein bottle amplitude \chapal\ and in particular the $\ZZ_3$
projector is complete.

\subsec{M\"obius amplitude}

The last one-loop amplitude to compute is the M\"obius amplitude
\eqn\chapaq{ \cM= c \, \int_0^\infty {dt\over t^4}\, 
                  {\rm Tr}_{open} \left[
                  {\Omega R\over 2}\, 
            \left( {1+\Theta+\Theta^2\over 3}\right)\, P_{GSO}\, 
                   e^{-2\pi t L_0} \right] ,}
where we also distinguish between untwisted and twisted contributions.
\smno
a.) {\it Untwisted sector}
\smno
Since the $7_1$ branes are reflected onto themselves under $R$, we expect
to find a nonzero contribution for $\cM_{7_1 7_1}$ with the $\Omega R$
insertion. Similarly, we expect that 
$\cM_{7_2 7_2}$ is only non-zero with the $\Omega R\Theta^2$ insertion and
that $\cM_{7_3 7_3}$ is only non-zero with the $\Omega R\Theta$ insertion.
The trace over the oscillator part is as described in \rgimpol. 
However, as in the annulus amplitude for the KK momenta in the 
$x_6$ and $x_8$ directions all $P=m/r$ contribute but  for
the winding in the $x_7$ and $x_9$ direction only doubly
wound modes $L=nr\sqrt{3}$ are invariant.
Thus, for the untwisted M\"obius amplitude we obtain
\eqn\chapar{ \cM_U=-c\, (1-1)\, \int_0^\infty {dt\over t^4}\, {M\over 4}\,
          {\th{0}{1/2}^4\, \th{1/2}{0}^4 \over
    \eta^{12}\, \th{0}{0}^4} \left( \sum_m e^{-2\pi t {m^2\over \rho}} 
     \right)^2
                   \left( \sum_n e^{-2\pi t {n^2 \rho 3}} \right)^2 }
with the argument exp$(-4\pi t)$.
\smno
b.) {\it Twisted sectors}
\smno
Here one realizes that the $(7_1 7_2)$ sector is invariant under
$\Omega R\Theta^2$, the $(7_2 7_3)$ sector is invariant under
$\Omega R$ and  the $(7_3 7_1)$ sector is invariant under
$\Omega R\Theta$. Thus, there are three contributions to the
twisted M\"obius amplitude. 
Using the general relation
\eqn\chapas{ { \th{a+1/2}{b} \over \th{a+1/2}{b+1/2}}(-q)=
            { \th{(a+1)/2}{a/2+b}\, \th{a/2}{(a+1)/2+b} \over
             \th{(a+1)/2}{(a+1)/2+b}\, \th{a/2}{a/2+b}  }(q^2) }
we obtain for the $\Theta$ twisted M\"obius amplitude
\eqn\chapat{\cM_\Theta=
            = -\kappa_{1} \,  c\, (1-1)\, \int_0^\infty {dt\over t^4}\, 
      {M\over 4}\, {\th{0}{1/2}^2\, \th{1/2}{0}^2\, \th{1/3}{1/2}\, 
        \th{-1/6}{0}\, \th{-1/3}{1/2}\, \th{1/6}{0} \over
        \eta^{6} \, \th{0}{0}^2\,  \th{-1/6}{1/2}\, \th{1/3}{0}\,
          \th{1/6}{1/2}\, \th{-1/3}{0} } }
and similarly for $\cM_{\Theta^2}$.
Transformation to tree channel again reveals the appearance of the 
complete projector for $\kappa_1=\kappa_2=1$.
\eqn\chapau{\eqalign{ \widetilde{\cM}=-c\, (1-1)\, &\int_0^\infty dl\, 
 {8M\over 3}\,  \Biggl[ {\th{1/2}{0}^4\, \th{0}{1/2}^4 \over
        \eta^{12}\, \th{0}{0}^4}
         \left( \sum_m e^{-\pi l {m^2 \rho 4}} \right)^2
                \left( \sum_n e^{-\pi l {4n^2 \over 3 \rho }} \right)^2 +\cr
     &\cr
      &3\, (\kappa_1+\kappa_2)\, 
          {\th{1/2}{0}^2\, \th{0}{1/2}^2\, \th{1/2}{1/3}\, 
        \th{0}{-1/6}\, \th{1/2}{-1/3}\, \th{0}{1/6} \over
        \eta^{6} \, \th{0}{0}^2\,  \th{1/2}{-1/6}\, \th{0}{1/3}\,
          \th{1/2}{1/6}\, \th{0}{-1/3} } \Biggr] \cr }}
with argument exp$(-8\pi l)$.

\subsec{Tadpole cancellation and massless spectrum}

Adding up all tree channel amplitudes from the Klein bottle, annulus 
and M\"obius strip we extract only one single  tadpole cancellation condition
\eqn\chapav{ {1\over 6}(M^2-16\,M+64)={1\over 6}(M-8)^2=0.}
Thus, in contrast to the 32 D-branes usually appearing in
pure $\Omega$ orientifolds, there are now only eight D7-branes of each kind. 
The next step is to compute the massless spectrum 
arising in both the closed and the open string sector and to show
that it indeed satisfies  anomaly cancellation.
 
The computation in the closed string sector differs from the analogous
computation in the case of the usual $\Omega$ orientifolds. 
In the untwisted sector we find the N=1 supergravity multiplet and
3 hypermultiplets. In a twisted sector we have nine $\ZZ_3$ fixed points. 
The one fixed point invariant under $\Omega R$ contributes 
1 hypermultiplet. The remaining  eight fixed points form four pairs
under the action of $\Omega R$ leading to 4 tensormultiplets and
4 hypermultiplets. Since there are two twisted sectors we end up with
the massless closed string spectrum
\smno
\cl{ 1 SUGRA + 8 tensormultiplets + 13 hypermultiplets .}
\smno
In the open string sector we find in the untwisted sector one
vectormultiplet in the adjoint of $SO(8)$ and one hypermultiplet in 
the adjoint, as well. Moreover, there appears one further
hypermultiplet in  the adjoint from the twisted sector, so that we
end up with 
\smno
\cl{ G=SO(8)  with 2 hypermultiplets in the adjoint .}
\smno
The closed and open string spectrum together satisfy the
anomaly constraint
\eqn\chapaw{ n_H-n_V +29\, n_T =273.}
It is not necessary to place all $7_1$ branes neither on top of each
other nor at the fixed line of $R$. The position of the $7_2$ and $7_3$
branes are of course still related to the position of the $7_1$ branes due
to the $\ZZ_3$ symmetry. 
Changing the position of the $7_1$
branes in the (68) plane is reflected in the effective theory as a Higgs 
branch on which the gauge symmetry is broken to some subgroups. For instance, 
placing four $7_1$ branes at a position $x_6=x_8=a$ with $a\ne 0$
and necessarily the  remaining four $7_1$ branes at 
$x_6=x_8=-a$, we derive the following massless spectrum
\smno
\cl{ G=U(4)\ \  with\ \  $2\times{\bf Adj} + 2\times {\bf 6}$ .}
\smno
This is related to the maximal spectrum via Higgsing and consistently 
satisfies the anomaly constraint, as well. 

We presented a purely perturbative treatment  of these
new permutational orientifolds thereby describing twisted open string
sectors by D7-branes intersecting at 60 and 120 degree angles. 
As we will see in the next sections some new aspects will arise when 
considering the other two possible $\ZZ_4$ and $\ZZ_6$ cases.

\newsec{The $\ZZ_4$ permutational orientifold }

In case of the $\ZZ_4$ orientifold some new issues arise which were
absent in the $\ZZ_3$ case. 
A first attempt to choose the rectangular $\ZZ_4$ symmetric 
lattice with basis vectors in the (67) and (89) direction, respectively, 
failed, as in the tree channel Klein bottle amplitude the complete 
projector $(1+\Theta+\Theta^2+\Theta^3)/4$ did not show up.
In particular in that case we found that the lattice part for
the $\{1,\Theta^2\}$  insertions in the Klein bottle trace differed 
from the result for the $\{\Theta,\Theta^3\}$  insertions.
The resolution to this puzzle is to start 
in the (89) plane with the lattice in figure 4
\fig{}{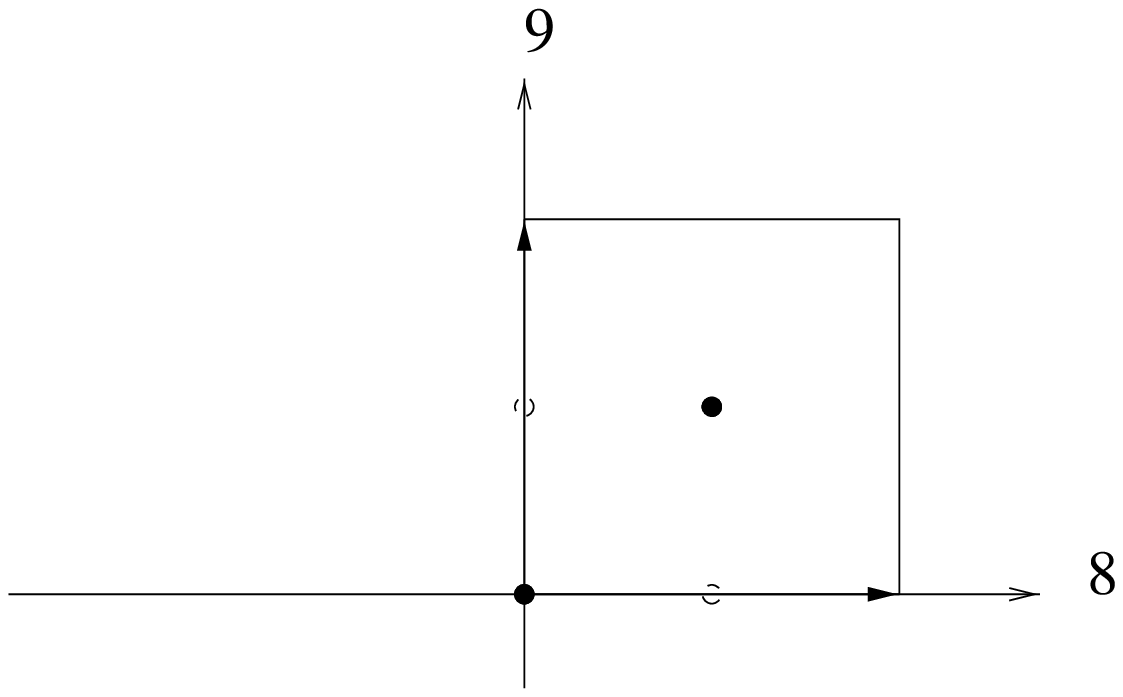}{7truecm}
\noindent
and in (67) plane with the rotated lattice shown  in figure 5.
\fig{}{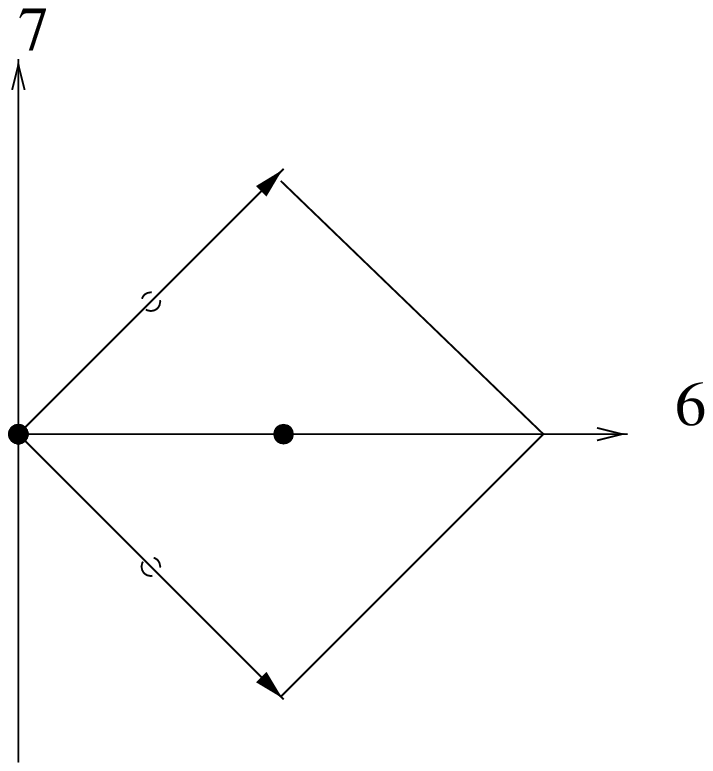}{4truecm}
\noindent
We have depicted the $\ZZ_4$ fixed points by black circles and
the additional $\ZZ_2$ fixed by white circles. 
From the figures it is evident that all four $\ZZ_4$ fixed points 
are also invariant under $R$ and that only eight of the sixteen $\ZZ_2$ fixed
points are invariant. Thus, the $\Theta$ and $\Theta^3$
twisted sector contribution to the Klein bottle amplitude is weighted
by a factor of four and the $\Theta^2$ twisted sector contribution
is weighted by a factor of eight.
The actual computation is completely similar to the $\ZZ_3$ case
and yields after all the following tree channel Klein bottle
amplitude
\eqn\chapba{\eqalign{ \tilde{\cK}=c\, (1-1)\, &\int_0^\infty dl\, 
 {2^6}\,  \Biggl[ {\th{1/2}{0}^4 \over
        \eta^{12}} \left( \sum_m e^{-\pi l {4 m^2 \rho}} \right)
                   \left( \sum_n e^{-\pi l {4 n^2\over \rho}} \right)
                   \left( \sum_m e^{-\pi l {2 m^2 \rho}} \right) \times \cr
                &\left( \sum_n e^{-\pi l {2n^2 \over  \rho }} \right)+
      2{\th{1/2}{0}^2\, \th{1/2}{1/4}\, \th{1/2}{-1/4} \over
        \eta^{6}\,   \th{1/2}{-1/4}\, \th{1/2}{1/4}} + \cr
      &2{\th{1/2}{0}^2\, \th{1/2}{-1/4}\, \th{1/2}{1/4} \over
        \eta^{6}\,   \th{1/2}{1/4}\, \th{1/2}{-1/4}} + 
        4{\th{1/2}{0}^2\, \th{1/2}{1/2}\, \th{1/2}{1/2} \over
        \eta^{6}\,   \th{1/2}{0}\, \th{1/2}{0}} \Biggr] \cr }}
with the argument exp$(-4\pi l)$. 
The last term in \chapba\ arises in the $\Theta^2$ twisted sector and
is actually zero. Nevertheless, 
its coefficient still demonstrates the appearance of 
the complete projector with numerical prefactors $4\sin^2(\pi k/N)$ in front
of the $\vartheta$-functions. 
The next step is to introduce appropriate D7-branes to cancel the tadpoles
in \chapba. The right choice to do that is shown in figure 6.
\fig{}{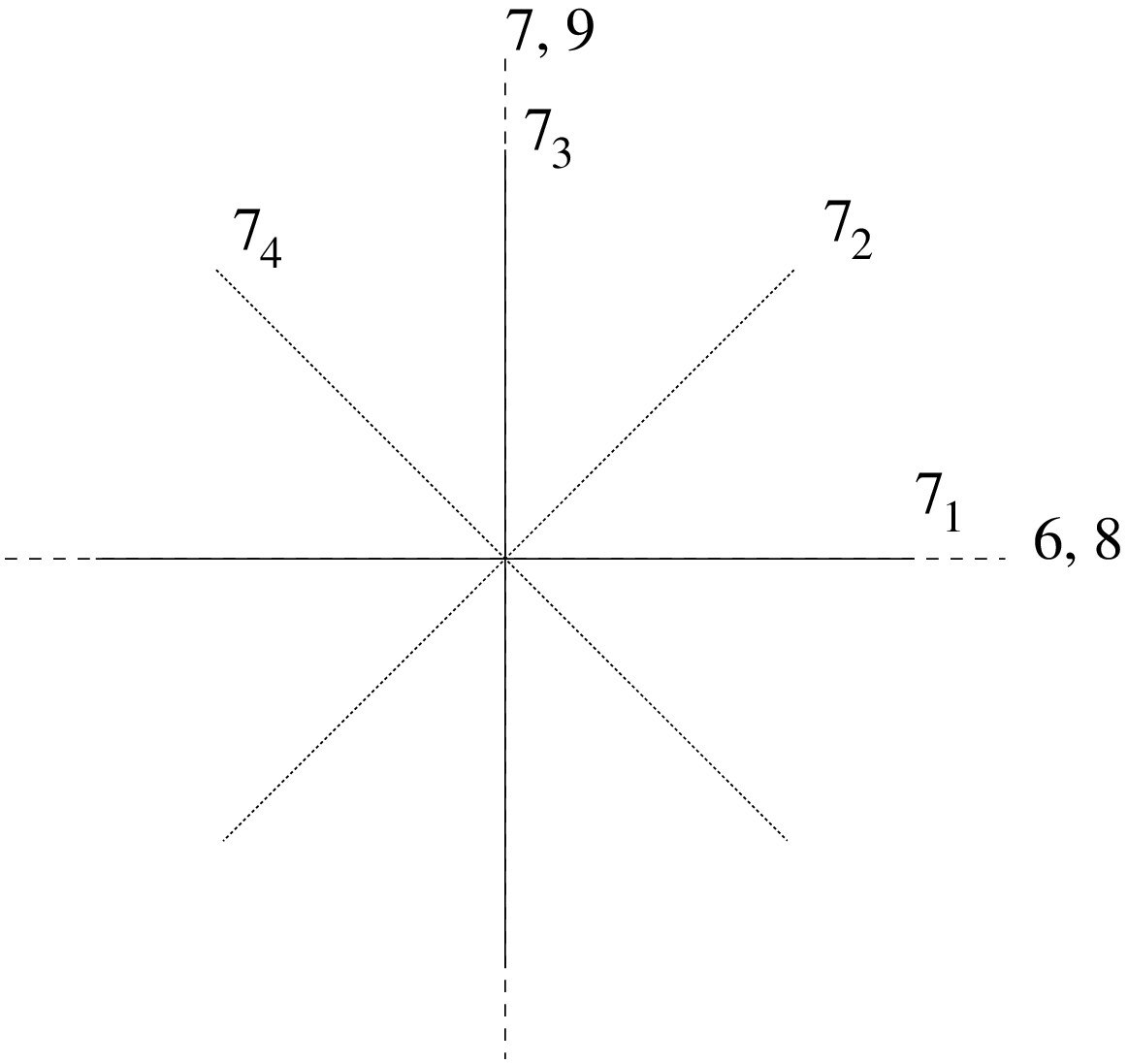}{7truecm}
\noindent
Unlike the $\ZZ_3$ case, here we have two sets of D7-branes which are not 
related by
$\ZZ_4$ rotations, namely the $\{7_1,7_3\}$ and the $\{7_2,7_4\}$ sets
close under $\ZZ_4$ rotations. Thus, eventually  we expect to get 
a product gauge group with two factors, 
one living on the first set of D7-branes and one living
on the second set of D7-branes.
Since there exists a $\ZZ_2$ subgroup leaving the different D7-branes
invariant we get an action of the $\ZZ_2$ on the Chan-Paton factors
of each D7-brane. As usual this action is described by a unitary matrix,
denoted as $\gamma_{\Theta^2}$. Due to the symmetry of the problem, we
assume in the following that these four a priori different $\gamma$-matrices
are all identical. The computation of the different terms in the
annulus amplitude is now straightforward. However, one finds that in the loop 
channel the complete projector does not constitute itself, in fact the
untwisted and $\Theta^{1,3}$ twisted sectors are fine, 
but the $\Theta^2$ sector
is too small by a factor of two. Looking into the details reveals
that in a $\ZZ_N$ orientifold the $\Theta^k$ twisted open string sector,
$A_{\Theta^k}$, has to be weighted by an extra factor of
\eqn\chapbb{  \kappa_k={\sin^2\left({\pi k\over N}\right)\over 
                        \sin^2\left({\pi \over N}\right) } .}
This is similar to ordinary orientifolds with a discrete 
background NSNS two-form B-field \refs{\rbfield,\rkakusix}, where also
the sector of open strings stretched between D9- and D5-branes is 
weighted by an extra factor of $2^{{\rm rg(B)}/2}$.  
Analogously, also in our case it turns out that these prefactors are  crucial 
for satisfying the anomaly constraint. 
Of course, it would be nice to have a deeper 
understanding of these extra factors from  the loop channel point of view.  
Unfortunately, we cannot offer a satisfying explanation  yet.
The complete tree channel annulus amplitude is
\eqn\chapbc{\eqalign{ \tilde{\cA}=c\, (1-1)\, &\int_0^\infty dl\, 
 {1\over 4}\,  \Biggl[ M^2\, {\th{1/2}{0}^4 \over
        \eta^{12}} \left( \sum_m e^{-\pi l {m^2 \rho}} \right)
                   \left( \sum_n e^{-\pi l { n^2\over \rho}} \right)
                   \left( \sum_m e^{-\pi l {2 m^2 \rho}} \right) \cdot \cr
                &\left( \sum_n e^{-\pi l { 2 n^2 \over  \rho }} \right)+
      8\, {\rm Tr}(\gamma_{\Theta^2})^2\, { \th{1/2}{0}^2\, \th{0}{0}^2 \over
           \eta^6 \th{0}{1/2}^2 } + \cr
         &2\, M^2\, {\th{1/2}{0}^2\, \th{1/2}{1/4}\, \th{1/2}{-1/4} \over
        \eta^{6}\,   \th{1/2}{-1/4}\, \th{1/2}{1/4}} +
         2\, {\rm Tr}(\gamma_{\Theta^2})^2\, 
          {\th{1/2}{0}^2\, \th{0}{1/4}\, \th{0}{-1/4} \over
        \eta^{6}\,   \th{0}{-1/4}\, \th{0}{1/4}} - \cr
   &4\, {\rm Tr}(\gamma_{\Theta^2})^2\, { \th{1/2}{0}^2\, \th{0}{1/2}^2 \over
           \eta^6 \th{0}{0}^2 } + \cr
       &2\, M^2\, {\th{1/2}{0}^2\, \th{1/2}{-1/4}\, \th{1/2}{+1/4} \over
        \eta^{6}\,   \th{1/2}{1/4}\, \th{1/2}{-1/4}} +
         2\, {\rm Tr}(\gamma_{\Theta^2})^2\, 
          {\th{1/2}{0}^2\, \th{0}{-1/4}\, \th{0}{1/4} \over
        \eta^{6}\,   \th{0}{1/4}\, \th{0}{-1/4}} \Biggr] \cr }}   
with the argument exp$(-4\pi l)$. 

In the untwisted M\"obius amplitude terms like 
\eqn\chapbd{ {\rm Tr}_{7_1 7_1}\left( {\Omega R\over 8}\, P_{GSO}\, 
 e^{-2\pi t L_0}  \right) }
give a non-zero result, whereas due terms like
\eqn\chapbe{ {\rm Tr}_{7_1 7_1}\left( {\Omega R\Theta^2\over 8} \,
          P_{GSO}\, e^{-2\pi t L_0} 
               \right) }
are zero and do not contribute due to the action on the oscillator modes. 
In the twisted sector the traces 
\eqn\chapbf{ {\rm Tr}_{7_1 7_3}\left( {\Omega R\over 2}
   {(\Theta+\Theta^3)\over 4} \, P_{GSO}\, e^{-2\pi t L_0} 
     \right)\quad {\rm and}\quad  
{\rm Tr}_{7_2 7_4}\left( {\Omega R\over 2}{(1+\Theta^4)\over 4}\,
      P_{GSO}\, e^{-2\pi t L_0}  \right) }
are non-vanishing. At first sight it might be confusing that the
$\ZZ_2$ twisted sectors contribute here, as in tree channel we expect 
to find some non-zero terms in the $\ZZ_4$ twisted sectors. However,
due to the different rules for the transformation 
from loop into tree channel in
the M\"obius amplitude and the relation \chapas, it finally comes
out just right. Moreover, the sectors in \chapbf\ are exactly those
weighted by this extra factor of two. This degeneration has to be taken 
into account
in the M\"obius amplitude, as well. As we shall find  in the $\ZZ_6$
example it can happen that the action of the $\ZZ_2$ subgroup must
be different in the degenerated sectors in order to guarantee 
completion of the projector in the tree channel M\"obius amplitude. 
For the tree channel M\"obius amplitude we find
\eqn\chapag{\eqalign{ \widetilde{\cM}=-c\, (1-1)\, &\int_0^\infty dl\, 
 {8\,M}\,  \Biggl[ {\th{1/2}{0}^4\, \th{0}{1/2}^4 \over
        \eta^{12}\, \th{0}{0}^4}
         \left( \sum_m e^{-\pi l {4m^2 \rho }} \right)
         \left( \sum_m e^{-\pi l {8m^2 \rho }} \right)
         \left( \sum_n e^{-\pi l {4n^2 \over  \rho }} \right) \cr
         &\left( \sum_n e^{-\pi l {2n^2 \over  \rho }} \right) 
      + (2\kappa_2)\,
          {\th{1/2}{0}^2\, \th{0}{1/2}^2\, \th{1/2}{1/4}\, 
        \th{0}{-1/4}\, \th{1/2}{-1/4}\, \th{0}{1/4} \over
        \eta^{6} \, \th{0}{0}^2\,  \th{1/2}{1/4}\, \th{0}{-1/4}\,
          \th{1/2}{-1/4}\, \th{0}{1/4} } \Biggr] \cr }}
with argument exp$(-8\pi l)$. Thus, also the M\"obius amplitude confirms
that the $(7_1 7_3)$ as well as the $(7_2 7_4)$ sector must be
weighted by an extra factor $\kappa_2=2$. Note, that the second term 
in \chapag\ is the sum of the $\Theta$ and the $\Theta^3$ twisted open 
string sector. 

Adding up all tree channel amplitudes we derive two tadpole cancellation 
conditions
\eqn\chapah{ {1\over 4}(M-16)^2=0,\quad\quad\quad 
            {\rm Tr}(\gamma_{\Theta^2})=0 ,}
yielding $M=16$ D7-branes of each type. The computation of the closed
string  massless spectrum gives one supergravity and 3 hypermultiplets
from the untwisted sector. In the twisted sector one has to be very careful
in treating the fixed points in the right way. The four $\ZZ_4$
fixed points contribute eight further hypermultiplets. The
eight $\ZZ_2$ fixed points invariant under $\Omega R$ give six
hypermultiplets and the remaining eight fixed points yield three
hypermultiplets and one tensormultiplet. Adding up all states we find
\smno
\cl{ 1 SUGRA + 1 tensormultiplets + 20 hypermultiplets .}
\smno
For the open strings  the untwisted sector gives a vectormultiplet 
in the adjoint of U(8)$\times$ U(8) and two hypermultiplets in the
antisymmetric representation for each gauge factor. The $\Theta$ and
$\Theta^3$ twisted sector contributes two  hypermultiplets in the
bifundamental  representation and finally the $\Theta^2$ 
twisted open string sector yields another two hypermultiplets in the
antisymmetric representation for each gauge factor. In the latter
we have taken into account the extra degeneration in this sector.
Summarizing  we get
\smno
\cl{ G=U(8)$\times$ U(8)\ \  with\ \    $4\times ({\bf 28},{\bf 1}) +
       4\times ({\bf 1},{\bf 28})+2\times ({\bf 8},{\bf 8})$. }
\smno
It is easily checked that the closed and open string spectrum 
above satisfies the anomaly constraint.

\newsec{The $\ZZ_6$ permutational orientifold }

In the $\ZZ_6$ orientifold there arises only
one new issue as compared to the $\ZZ_3$ and $\ZZ_4$ examples to be
discussed shortly. The rest of the computation is completely analogous
and we will skip most of the details.
Analogous to what we encountered in the $\ZZ_4$ case,
we are free to choose in the  (89) plane the lattice in figure 7.
\fig{}{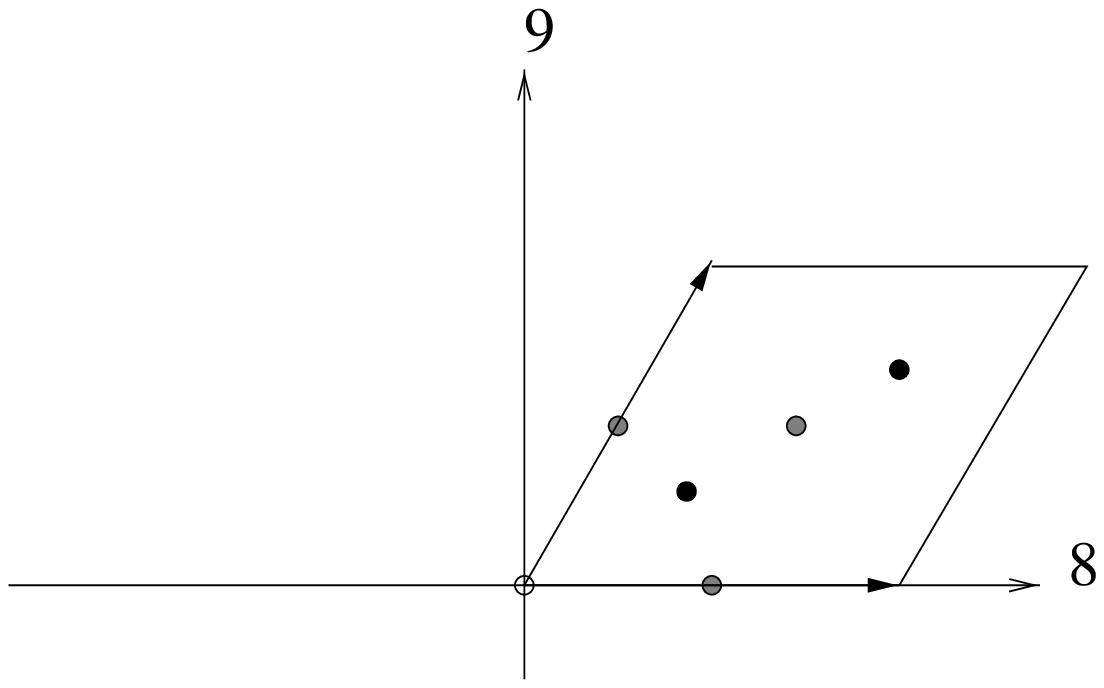}{7truecm}
\noindent
In the (67) plane we are forced to take the rotated lattice as shown
in  figure 8.
\fig{}{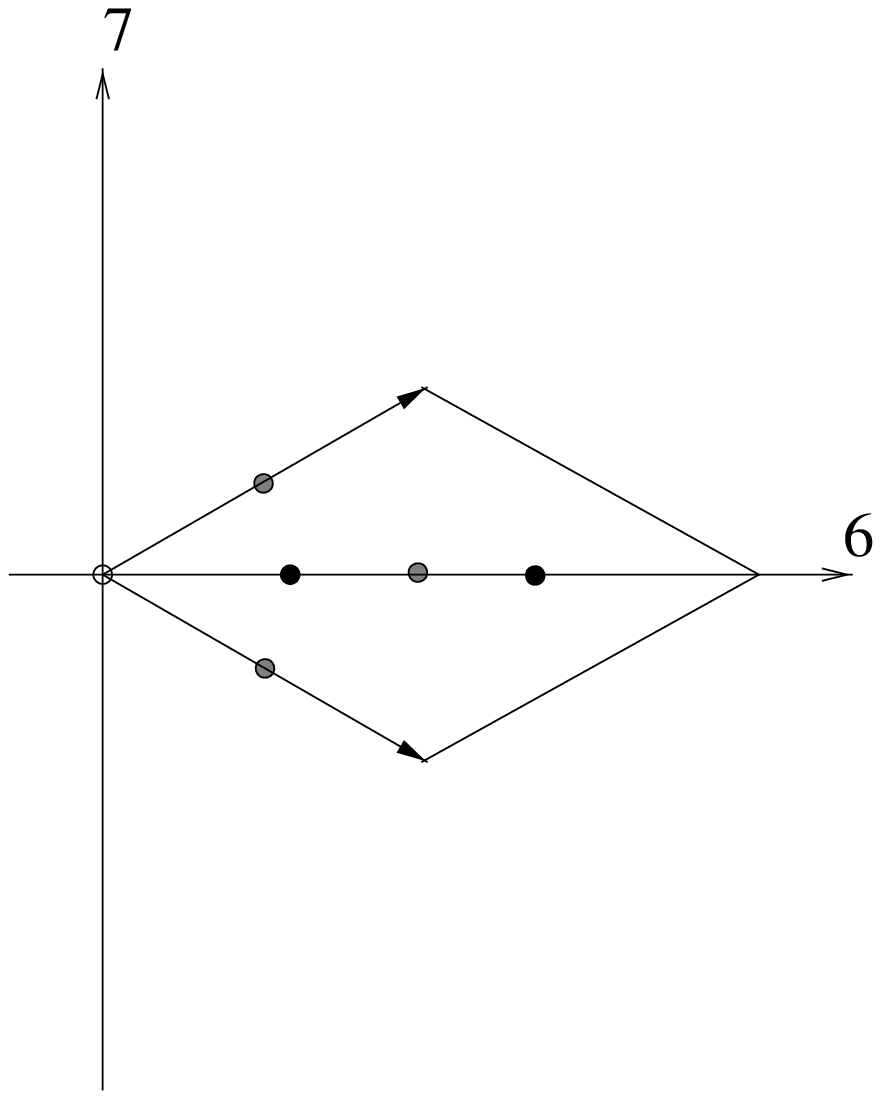}{5truecm}
\noindent
The one $\ZZ_6$ fixed point is denoted by a white circle, the additional
two $\ZZ_3$ fixed points
by black circles and the additional three $\ZZ_2$ fixed points by  grey
circles. The behaviour of the fixed points under the reflection $R$ is 
obvious, namely three of the nine $\ZZ_3$ fixed points and four of the 
sixteen $\ZZ_2$ fixed points are invariant under
$R$. Computing the Klein bottle amplitude we realize that these numbers are 
exactly those needed to get the complete $\ZZ_6$ projector
in tree-channel. The tadpoles appearing  in the Klein bottle amplitude
can be cancelled by introducing six different D7-branes
rotated against each other by thirty degrees as shown in figure 9. 
\fig{}{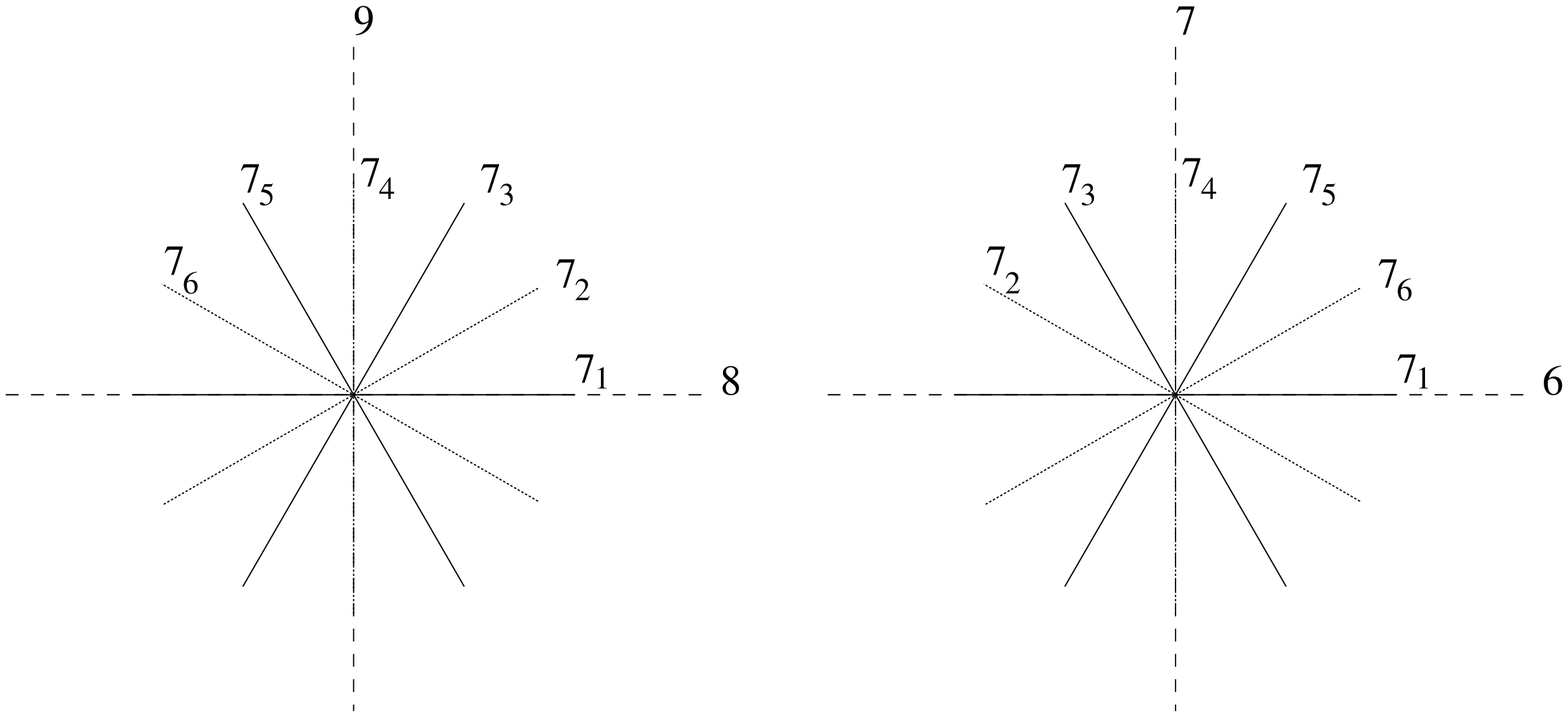}{16truecm}
\noindent
Under the action of $\ZZ_6$ the elements of two sets $\{7_1,7_3,7_5\}$ and 
$\{7_2,7_4,7_6\}$ are transformed among themselves. 
Open strings of type $(7_i 7_i)$ can be understood as the untwisted sector
and open strings of type $(7_i 7_{i+3})$ as the $\Theta^3$ twisted sector.
The remaining two $\Theta^{2,4}$ and $\Theta^{1,5}$ twisted sectors 
are given by
open strings of type $(7_i 7_{i+2})$ and $(7_i 7_{i+1})$, respectively.
In the computation of the twisted sector annulus amplitude, 
we have to introduce those extra factors \chapbb.
The resulting expressions are  very similar to the $\ZZ_4$ annulus
amplitude \chapbc. 

The only sectors contributing to the loop channel M\"obius amplitude
are the untwisted one and the one arising from open strings
in the $(7_i 7_{i+2})$ sector. In the $(7_1 7_3)$ sector, for instance, 
the two terms 
\eqn\chapca{ {\rm Tr}_{7_1 7_3}\left( {\Omega R\over 2}
 {\Theta+\Theta^4\over 6}
\, P_{GSO}\, e^{-2\pi t L_0} \right)}
are non-vanishing. The first one gives rise to a contribution of the
$\Theta$ twisted sector in the M\"obius tree channel amplitude and
the second one yields a contribution to the $\Theta^2$ twisted sector.
The projector tells us that the relative normalization of these two 
factors in the tree channel
must be $\widetilde{\cM}_{\Theta^2}=3\widetilde{\cM}_{\Theta}$. However, from
the annulus amplitude the overall normalization of the  
$(7_i 7_{i+2})$ sector is fixed to be three. It is possible 
to repare that by
requiring that in the three degenerated $(7_i 7_{i+2})$ sectors the action
of $\Omega R$ is not identical. One sector must be invariant and the other
two must be exchanged under the action of $\Omega R$.  
Fortunately, our guiding principle again
allows us to fix everything.
Apparently, the different action of $\Omega R$ on 
the three degenerated sectors
has dramatic consequences for  the computation of the massless spectrum.

The two tadpole cancellation conditions in the $\ZZ_6$ case are
\eqn\chapah{ {1\over 2}(M-8)^2=0,\quad\quad\quad 
            {\rm Tr}(\gamma_{\Theta^3})=0 .}
The computation of the closed
string  massless spectrum yields  one supergravity and 3 hypermultiplets
from the untwisted sector. 
Two further hypermultiplets come from the  $\ZZ_6$
fixed point. The
nine $\ZZ_3$ fixed points give rise to two tensormultiplets and
eight hypermultiplets. Finally, the sixteen $\ZZ_2$ fixed points 
yield one tensormultiplet and five hypermultiplets.
Adding up all these states we find
\smno
\cl{ 1 SUGRA + 3 tensormultiplets + 18 hypermultiplets .}
\smno
For the open strings  the untwisted sector gives rise to a vectormultiplet 
in the adjoint of U(4)$\times$U(4) and two hypermultiplets in the
antisymmetric representation for each gauge factor. The $\Theta$ and
$\Theta^5$ twisted sector contribute two  hypermultiplets in the
bifundamental  representation.  Taking into account the above mentioned 
subtlety in the $\Theta^{2,4}$ 
twisted open string sector we obtain  four hypermultiplets in the
antisymmetric representation and one hypermultiplet in the
adjoint representation for each gauge factor.
Finally, four further hypermultiplets in the
bifundamental  representation arise in the $\Theta^3$ twisted open 
string sector.
Summarizing  we get
\smno
\cl{ G=U(4)$\times$U(4)\ \   with\ \  $6\times ({\bf 6},{\bf 1}) +
       6\times ({\bf 1},{\bf 6})+({\bf Adj},{\bf 1}) +
        ({\bf 1},{\bf Adj}) + 6\times ({\bf 4},{\bf 4})$, }
which together with the closed string spectrum indeed satisfies the anomaly 
constraint.

\newsec{Conclusion}

We have studied a new class  of orientifolds in which 
the world-sheet parity transformation is combined with a permutation
of the internal coordinates. 
As the main new issue we encountered non-zero contributions to the
twisted sector Klein bottle amplitude, which fortunately can be cancelled 
by introducing D7-branes intersecting at non-trivial angles into
the background. We have studied all possible N=1 supersymmetric
examples in six-dimensions,
performed the complete tadpole cancellation computation and found after
all  reasonable anomaly-free massless spectra. We would like to emphasize 
again that the main tool or guiding principle
used in the construction  was the requirement of always getting
the complete $\ZZ_N$ projector in the tree channel amplitude after
modular transformation of the loop channel. 

There exist a lot of open questions related to the permutational
orientifolds  we have presented
in this paper. For instance, it is interesting to study models of this type
in four space dimensions with N=1 supersymmetry \rfut. Moreover, these
models are particularly suitable to yield non-tachyonic Type 0B 
generalizations \rfutb. It is not evident what happens when one turns
on some background two form flux and finally there might exist some dual
F-theory descriptions.   
\bigno\bigno

\centerline{{\bf Acknowledgements}}\pano
We would like to thank Dieter L\"ust for encouraging discussions.
\bigno

\listrefs

\bye